\documentclass[12pt]{article}
\usepackage{graphics,epsfig}
\begin{document}
\title{The September 11 Attack: \\A Percolation of Individual Passive Support}
\author{Serge Galam
\\Laboratoire des Milieux D\'esordonn\'es et H\'et\'erog\`enes, \\
Tour 13, Case 86, 4 place Jussieu, 75252 Paris Cedex 05, }
\date{(galam@ccr.jussieu.fr)}
\maketitle

\begin{abstract}

A model for terrorism is presented using the theory of percolation. Terrorism power is
related to the spontaneous formation of random backbones of people who are sympathetic to
terrorism but without being directly involved in it. They just don't oppose in case they
could. In the past such friendly-to-terrorism backbones have been always existing
but were of finite size and localized to a given geographical area. The
September 11 terrorist attack on the US has revealed for the first time the existence of a
world wide spread extension. It is argued to have result from a sudden world percolation
of otherwise unconnected and dormant world spread backbones of passive supporters.
The associated strategic question is then to
determine if collecting ground information could have predict and thus avoid such a
transition. Our results show the answer is no, voiding the major criticism against
intelligence services. To conclude the impact of military action is discussed.
\end{abstract}


\newpage
The recent September 11 terrorist attack on the US came as a total and dramatic blow to 
all experts on terrorism, intelligence services and military hierarchy. Here, terrorism
designates the use of random violence against civilians in the purpose to kill them.
While the problem is extremely complex, complicated and difficult, a different view from
the physics of disorder may be useful in shedding some new light on it.

In the past years
physicists have been dealing with social and political behaviors using some concepts and
tools from Statistical Physics ~\cite{monde,frank,helb,sex,barth,nadal}. Here we are using
the theory of percolation ~\cite{perco,pajot,perco-sor} to analyze the connection between
terrorism activity and the surrounding population attitude. We are neither investigating
the terrorist net itself nor its internal mechanisms. 

Our work does not aim at an exact description of terrorism complexity. Making some crude 
approximations it allows exhibiting an essential characteristics of terrorism by linking
its capacity of destruction to the surrounding population attitude. In particular a target
is set to be reachable once it is located within an area covered by a cluster of people
who are passively consenting to the terrorist cause. The September 11 terrorist attack on
the US is given an explanation in terms of the first worldwide percolation of such a
cluster of passively consenting people. In parallel collecting ground information has
proven unable to assess the associated current world level of related terrorism threat.
Military action also appears of no use against it.

Passive supporters are normal people who do
not need to express explicitly their position. It is a dormant attitude that results
from an individual opinion. They are unnoticeable. They just do not oppose a terrorist
act in case they could. They are sharing independently an identical opinion of
identifying with the terrorist cause. They do not need to communicate between them.
Mainly concentrated within the  terrorist home area they are randomly spread in the
whole population. We analyse their distribution in terms of percolation theory. 

To make our model more explicit, we assume the world is a grid within a
continuum percolation picture. Sites are randomly distributed on a plane and when
separated by a distance less than some maximum length, they are regarded as
nearest neighbors. To move on the grid
requires to go through a continuous path of neighboring sites. People
are distributed randomly over all the sites. Therefore for someone to change site, it
has to simultaneously exchange sites with a nearest neigbhor on the grid making its
agreement a prerequisite for its one site move. Accordingly for a terrorist to go
from one home site to a target site requires to find a continuous path of passive
supporters with whom it can switch successively. All of them do not need to cooperate
collectively. The interaction is pairwise, local and restricted to the moving
terrorist. Only one passive supporter is involved at a time. In contrast a non passive
supporter does not allow the site exchange thus blockading the terrorist motion.

The random distribution of passive supporters produces random clusters of neigbhoring
passive supporters. No one is aware it belongs to a definite cluster. Such a cluster
existence comes to life only via the moving of a terrorist on it.
Otherwise it is totally virtual. However it is both the size and the localization of
such virtual clusters which determine the global level of terrorist threat as well as
its in reach potential targets. Up to now the passive supporters of a terrorist cause
have been always mainly concentrated with the geographical area of the terrorist home.
It usually give rise to one major home cluster with sometimes few additional clusters
further away but unconnected to the home one (see Fig. 1). It explains why all known
terrorism has been always geographically anchored to finite size areas like for instance
recently in Ireland, Corsica or Euskadi.

\begin{figure}
\begin{center}
\centerline{\epsfbox{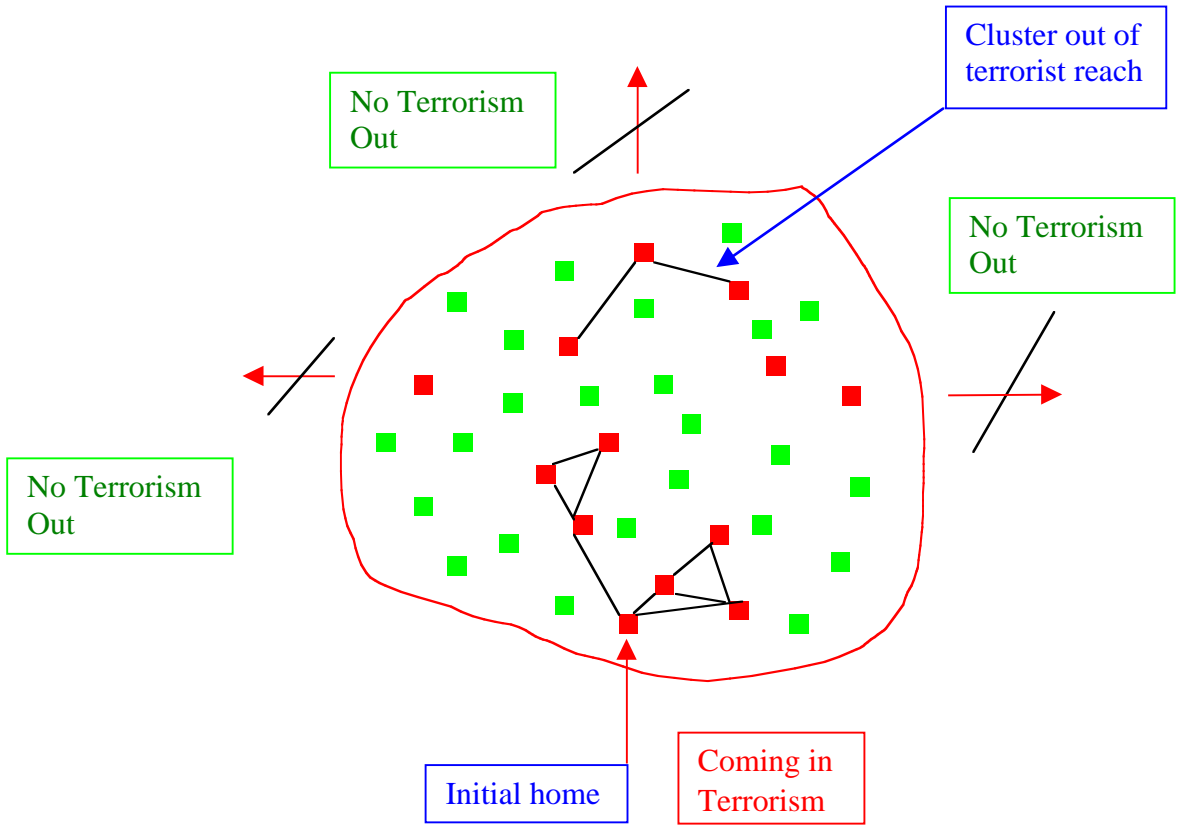}}
\caption{Black squares are open to terrorism. Grey ones are closed to it. The connected
black squares are clusters open to terrorism. The larger one below is being used by
terrorists making all its sites potential targets. The smaller cluster above is of no use
to terrorist since unconnected to their in coming source. Moreover the initial home
incoming terrorism is trapped within the country without possibility to reach another
neighboring country.}
\end{center}
\end{figure}    

However one of the characteristics of current terrorism has been its capacity in creating 
many passive supporters spread all over the world. It also succeeded
in producing one huge home cluster covering in part Afghanistan (see Fig. 2). Nevertheless
for years other existing world backbones stayed out of reach to its activity making them
invisible and unused. Accordingly the past years of continuing dynamics of converting more
and more people to the terrorism cause went unnoticeable. At least, it did not rise any
concern. After all it was just a question of opinion spreading. And indeed, since all new
and enlarged backbone of passive supporters were staying unconnected to the home one, they
were of no practical significance on terrorism activity. Their size 
increase went without any consequence.

\begin{figure}
\begin{center}
\centerline{\epsfbox{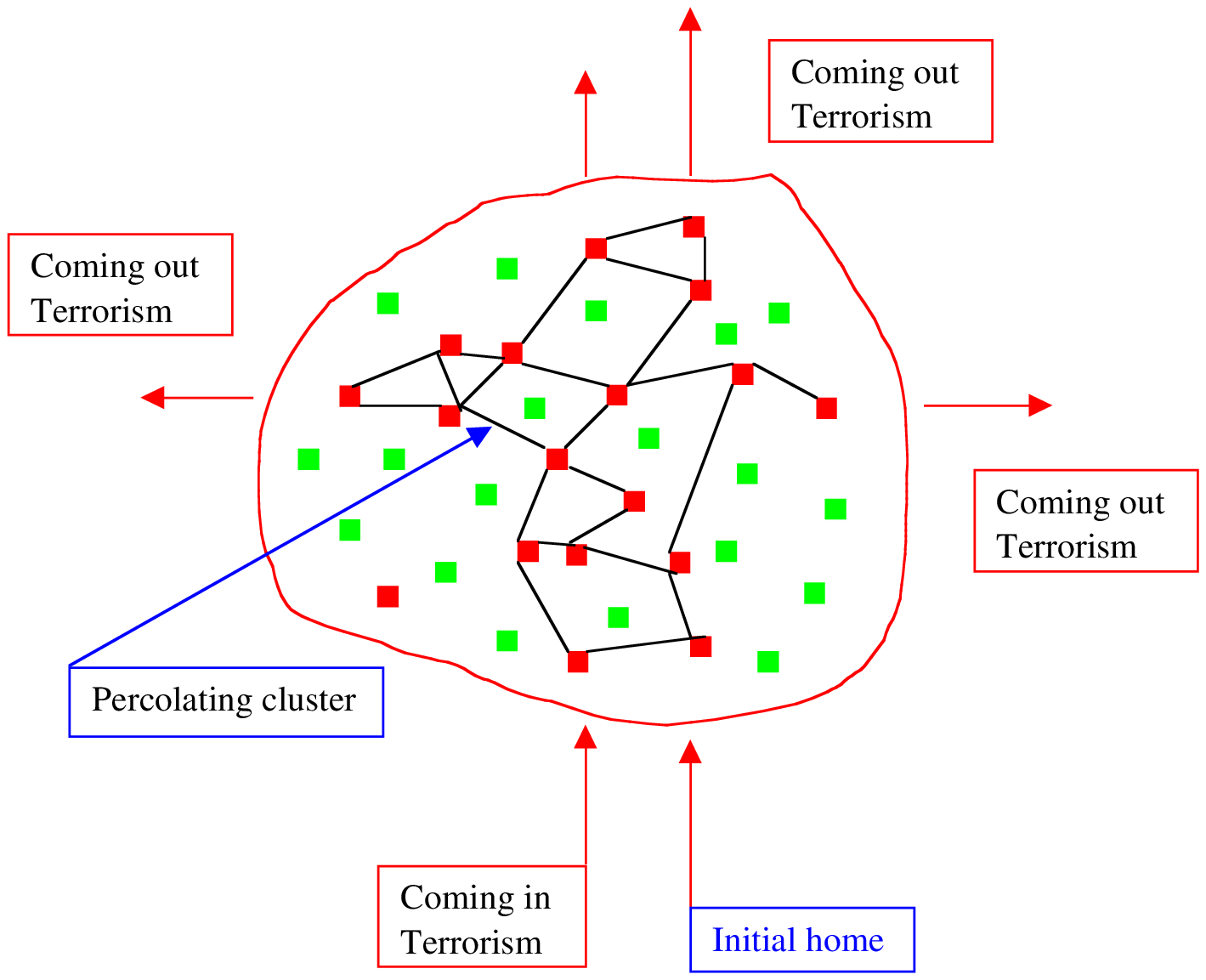}}
\caption{Black connected squares are a percolating cluster. Incoming terrorism can get out
towards any other neighboring country.}
\end{center}
\end{figure}

However at some density coverage of passive supporters a drastic geometric phenomenon
occurs. All of a sudden and at once many
existing spread backbones got connected to merge into one unique huge world wide
cluster. It must have been the first passive social percolation phenomenon in terrorism
history. Passive means the phenomenon occurs without any particular organized social
behavior. It is a geometrical result that does not require interactions among its
ingredients. The associated level of threat became automatically infinite. We argue the
September 11 terrorist attack on the US is the immediate result of that percolating
cluster. It is worth to stress at this stage that the few previous long-ranged terrorist
attacks, like the one by the Japanese Red Army Group at Tel Aviv airport three decade ago,
are associated to a dangling path rather than a percolating cluster.

Such an explanation rises the question on whether it was possible to predict this
world wide percolation. To answer such a strategic question and keep the
presentation simple we set black a site open to terrorism move and grey a blockading one. 
To illustrate the demonstration the percolation threshold is supposed to be at fifty
percent of the whole grid sites. The qualitative results do not depend on
this choice. Moreover within the qualitative level of the present paper it may be
sufficient to state that approximately a structure is connected if the majority of its
substructures are connected. 
Accordingly as long as less than fifty percent of the world sites are black
there exists no world percolating cluster and therefore no world level to
terrorism threat.

Now, at this stage we are aiming at measuring a world state which
result from individuals attitudes. Therefore to find the distribution of world passive
supporters requires getting ground information about individual attitudes. One
natural approach to achieve such a goal is to use ground people to report on what they
see. Indeed let us examine such a scheme.

In principle each person sees things which it could
report on. However these reports would score up to billions making it impossible to
collect all of them, even with very large resources. The construction of a representative
sample of such an infinite ensemble is then a prerequisite. Let us assume it exists.
Along with, every person of the sample reports on the black or grey occupation of
the various sites it can see. The associated grid coverage thus defines an individual
area. The goal is then to state if it is an area open to terrorist move or not. To do so
requires to aggregate and synthezise the resulting color of several sites which are
respectively either black or grey. Being concerned with the percolation of the whole
world grid, a natural way to proceed is to apply the same above hypothesis of fifty
percent at the level of individual areas. A given area with more than fifty percent of
black sites is black, otherwhise it is grey.

Once this area step is completed, the next one is to collect and treat all available
world spread colored areas to determine the world status with respect to its passivity to
terrorism. But here come a genuine difficulty. Real information is not just a black or
grey color. Indeed, the more from the ground an information is, the more specific to the
ground it is. Cultural biases, religion, poverty, military pressure and many
other local characteristics dress heavily any ground information. In term of information
theory the signals are very noisy.
Accordingly all area informations cannot be just added together. They must be
grouped by families such that their respective dispersions are not too
large to be understandable by one unique person which has to make a clear
synthesis report. For individual areas, a clan can be such a natural frame to a
synthesis.

Then, keeping our fifty percent criterium, a color clan is determined 
using a simple majority rule among the various colors of its associated individual areas.
A grey clan is hostile to terrorism while a black clan don't oppose it.
Along the same aggregating-synthezising process, clans have to be grouped by ethnies,
ethnies by provinces, provinces by states, states by countries, countries by
continent and continents gives the world. At each iterative step, colors are
respectively determined using the same local majority rule ~\cite{poli}.
It means, at each step
we are increasing the size of territories for which we are setting the color. For instance
starting from individual areas black and grey we go to each associated clan area which
becomes either black or gray. In other words a posteriori we are turning all individual
areas to the same and unique color of its corresponding clan (see Fig. 3). And so forth
going up to higher levels.

\begin{figure}
\begin{center}
\centerline{\epsfbox{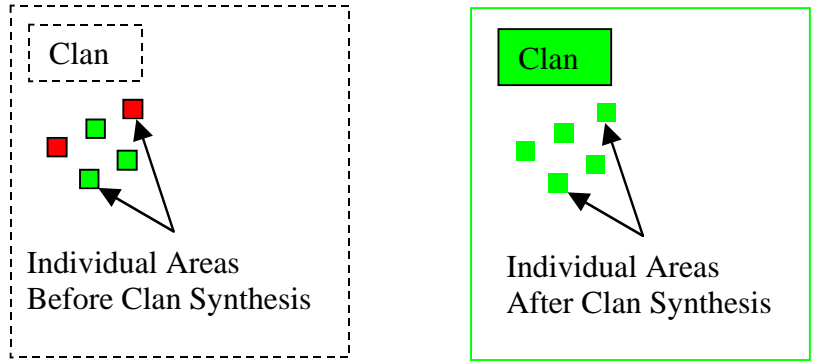}}
\caption{Synthesizing process. (i) On the left side 5 individual
areas before the clan level synthesis with 2 black and 3 grey areas. On the right side
after the clan level synthesis which makes the clan grey. Now all inside 5 areas are looked
upon as grey areas.}
\end{center}
\end{figure}

To proceed with a quantitative scheme, we note $n=0$ the ground site level of aggreagtion,
$n=1$ the individual areas,
$n=2$ the clans, $n=3$ the ethnies, $n=4$ the provinces, $n=5$ the states, $n=6$ the
countries,
$n=7$ the continents and $n=8$ the World. At each level a black color means the level
allows terrorist activity on the corresponding territory. Green means it forbids it (see
Fig. 4). For instance one black country surrounded by grey ones confines possible 
terrorism
activities within this country. Such a possibility gives light to the fact that
several countries stated that prior to September the 11 they did warn the US about the
possibility of some terrorist action. But such a warning could not be credited since the
US would have to be sure the world was percolating to engage a preventive military action.
Otherwise any military move would have been rejected by the whole world as arbitrary
aggressive.

Let us now illustrate the quantitative mechanism at work in the model by the
repeated color rescaling from site colors to the world color. To solve simple
equations we choose arbitrary numbers for each grouping step. Exact numbers could be
used but it would only make the equations more fastidious without changing the
main qualitative result. We take 5 sites per individual area, 5 individual areas per clan,
5 clans per ethny, 5 ethnies per province, 5 provinces per state, 5 states per
country, 5 countries per continent and 5 continents to the world. It implies  5 $\times$
5 $\times$ 5 $\times$ 5 $\times$ 5 $\times$ 5 $\times$ 5 = 78 125 human personals on the
ground. Each one of them reports on 5 different sites. 

Assuming everything is randomly distributed in our collecting system we can write one
equation giving the probability $p_n$ to have a black entity at level $n$ as
function of the probability $p_{n-1}$ to have one at the level below ${n-1}$, 
\begin{equation}
p_n = p_{n-1}^5+5 p_{n-1}^4 (1-p_{n-1})+10 p_{n-1}^3 (1-p_{n-1})^2 \ ,
\end{equation}
where $n$ runs from 1 for individual areas to 8 for the world. The value $p_0$
gives the density of reported black ground sites. Suppose $p_0=0.47$, the
percolation transition is about to happen. We can then make our intelligence
collecting ground information system at work to check its validity. 

Iterating Eq. (1) from $p_0=0.20$ gives $p_1=0.06$ with
$p_2=p_3=p_4=p_5=p_6=p_7=p_8=0.00$ where 0.00 symbolizes fractions below half a percent.
The collecting procees makes $20\%$ of black sites a small ground noise not to worry
about. An increase of black support to $30\%$ results in the series $p_1=0.16$, $p_2=0.03$
and $p_3=p_4=p_5=p_6=p_7=p_8=0.00$. Reaching
$40\%$ leads to $p_1=0.32$, $p_2=0.19$, $p_3=0.05$ and again
$p_4=p_5=p_6=p_7=p_8=0.00$. Getting close to the transition point at $p_0=0.47$ gives
$p_1=0.44$, $p_2=0.40$, $p_3=0.31$, $p_4=0.18$, $p_5=0.04$, $p_6=0.00$, $p_7=0.00$ and
$p_8=0.00$. If some indication is emerging about something going on within countries
with $p_5=0.04$, at the country level noone is a threat. The
aggregation procees has dropped out the existing
$47\%$ of ground black sites. It would lead to a total
disaster in term of any forecast. At the same time from $p_0=0.55$ the aggregation
process yields $p_1=0.59$, $p_2=0.67$, $p_3=0.80$, $p_4=0.94$ and now
$p_5=p_6=p_7=p_8=1$. The conclusion would be a total hysterical view of the whole
world in the terrorist camp.

\begin{figure}
\begin{center}
\centerline{\epsfbox{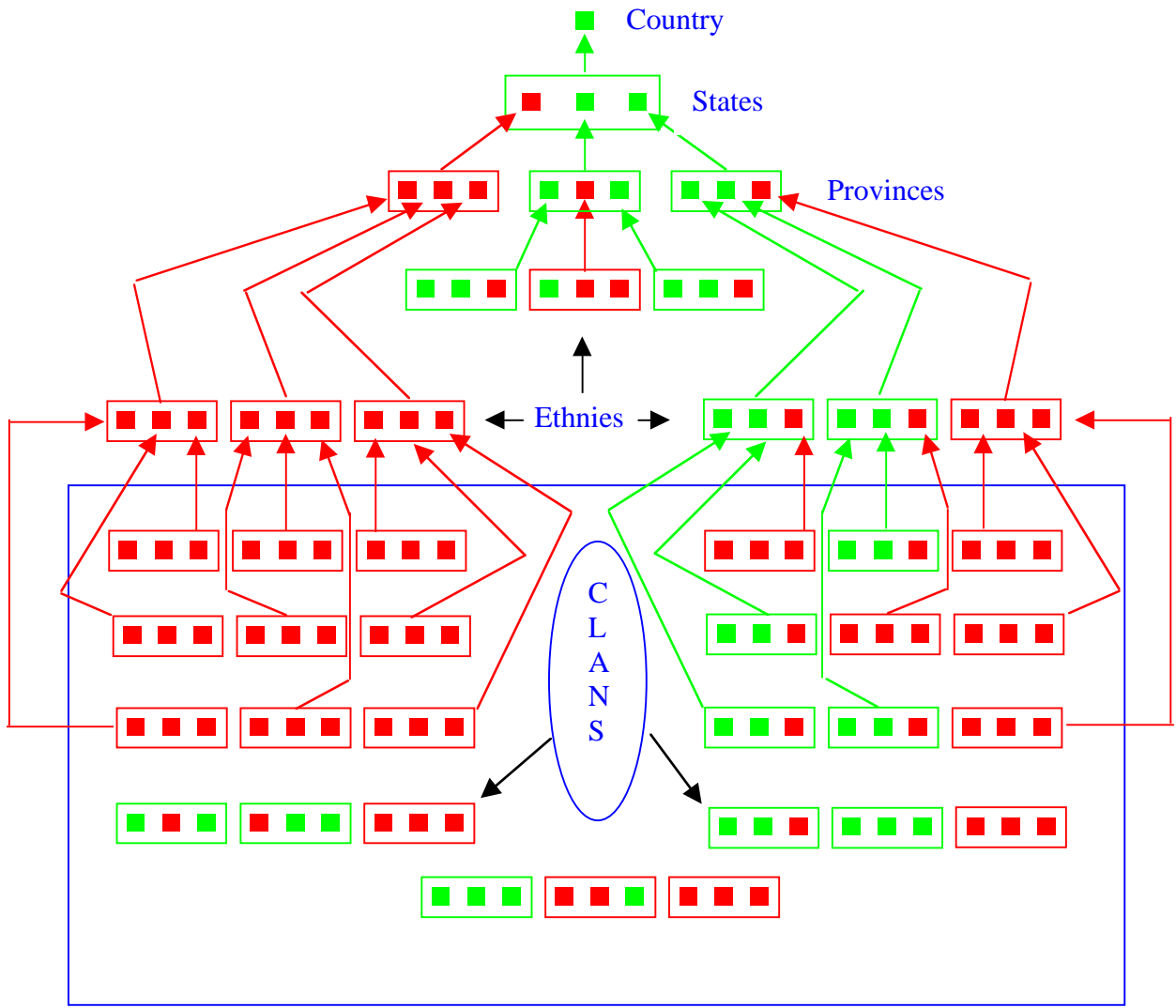}}
\caption{A particular illustration for one country with 3 states, 9, provinces, 27
ethnies and 72 clans. Although 48 clans are black against 24 grey, the intelligence
machine yields a grey country due to the peculiar distribution of the grey clans.
Below 27 clans without outgoing arrows are connected to the above 9 ethnies without
incoming arrows.  }
\end{center}
\end{figure}

From above numbers, the intelligence machine has proven its ability to determine
exactly the current world color but it misses totally the corresponding ponderation of
the leading color. Moreover the vicinity of a possible global shift is competely
missed. When the shift to black occurrs it is too late to react. At this stage the world
appears either ``good" or ``bad".
And indeed the performance is even worse. In many real
life situations, it may happen a synthesis cannot provide a clear
characterization on whether an area is black or grey. In that case to avoid a dramatic
mistake a doubting synthesis is associated to a grey signal.
No one would decide military action unless the evidences are unquestionable.

To implement this zero mistake requirement within our
model we introduce some even size collecting groups which may exhibit
equal number of black and grey sites. In this case the color is grey. To illustrate this possibility let us go from 5 to 6 in above intelligence
machine with 6 $\times$ 6 $\times$ 6 $\times$ 6 $\times$ 6 $\times$ 6 $\times$ 6 = 279
936. Eq. (1) becomes,
\begin{equation}
p_n = p_{n-1}^6+6 p_{n-1}^5 (1-p_{n-1})+15 p_{n-1}^4 (1-p_{n-1})^2 \ .
\end{equation}

Again iterating Eq. (2) from $p_0=0.47$ gives respectively $p_1=0.29$,
$p_2=0.06$, and $p_3=p_4=p_5=p_6=p_7=p_8=0.00$. Although there exist $47\%$ of black
sites the machine gives every level grey already at the clan one. Checking on
an initial $p_0=0.60$ leads to $p_1=0.54$,
$p_2=0.43$, $p_3=0.22$, $p_4=0.02$, and $p_5=p_6=p_7=p_8=0.00$. The result is dramatic
with a total failure in determining even the current state of the world which is much
above the percolation threshold.
Changing above numbers does not modify the qualitive result.

To conclude the use of ground information can at best determine the current dominant 
ground state but cannot assess the associated level of potential threat, in particular
the vicinity of a phase transition. And most likely, since a large-scale military action
cannot be undertaken with any doubt, it even misses totally a ground shift to a state of
maximum world danger. On this basis our results make void the major criticism against
intelligence services that would have failed in opposing current terrorism due to its
neglecting of human ground personal. The hint to a better efficiency of intelligence
services should to be looked upon other direction than coming back to old ground
intelligence practice.

Last but not least our finding demonstrate that the use of military power cannot reduce 
the level of threat attached to the existence of the percolating backbone of passive
supporters. Indeed to suppress a percolating cluster requires getting down the number of
passive
supporters below the percolation threshold. Imagine there exist 54 percent of passive
supporters within a
population of 100 millions. A military action would aim at neutralizing say 10 percent of
them. To achieve such a goal would require the neutralization of more than 20 percent of
the whole population (20 millions) whose moreover half are innocents since passive
supporters are
randomly distributed and unnoticeable. It is morally unacceptable though achievable with
currently available weapons of mass destruction. \\

\textbf{Acknowledgement}\\

I would like to thank D. Makinson and the anonymous referee for fruitful comment on the
subject.

\end{document}